\documentstyle[psfig,namedreferences]{kluwer}

\newcommand{\eg}{{\it eg.}}
\newcommand{\et}{{\it et~al.}}

\newcommand{\CGRO}{{\it CGRO}}
\newcommand{\EGRET}{{\it EGRET}}
\newcommand{\GLAST}{{\it GLAST}}
\newcommand{\COSB}{{\it COS~B}}

\newcommand{\Gam}{$\gamma$~rays}
\newcommand{\gam}{$\gamma$-ray}
\newcommand{\pair}{$e^+/e^-$}
\newcommand{\XX}{\tau}

\newcommand{\psip}{\psi}
\newcommand{\psim}{\psi^\star}
\newcommand{\psiv}{\psi^{\rm nv}}
\newcommand{\lamp}{\lambda}

\newcommand{\Lamp}{\Lambda}
\newcommand{\Lamm}{\Lambda^\star}
\newcommand{\Lamv}{\Lambda^{\rm nv}}

\begin{opening}
\title{Sensitivity of $\gamma$-Ray Detectors to Polarization}
\author{I.-A. \surname{Yadigaroglu}}
\institute{Department of Physics, Stanford University, Stanford, CA 94305-4060}
\runningtitle{GAMMA-RAY DETECTORS}
\runningauthor{I.-A. YADIGAROGLU}
\end{opening}

\begin{document}

\begin{abstract}
Previous studies have shown that the largest \gam\ detector to date,
\EGRET, does not have useful polarization sensitivity. We have explored
here some improved approaches to analyzing \gam\ pair production events,
leading to important gains in sensitivity to polarization. The performance
of the next generation \gam\ instrument \GLAST\ is investigated using a
detailed Monte Carlo simulation of the complete detector.
\end{abstract}

\section{Introduction}

The \EGRET\ instrument aboard \CGRO\ is the largest \gam\ detector to
date and has collected several thousand photons above 30 MeV for the
strongest sources, the Vela and Crab pulsars. Most of the energy
output of young pulsars is in beamed \Gam\ which emission models
predict to posses a high degree of linear polarization (\eg\ Romani
and Yadigaroglu 1995), and polarization observations would prove a
particularly potent probe of magnetospheric geometry and physics.

At \gam\ energies above a few MeV, photon interactions with matter are
increasingly dominated by pair production. Detectors in this energy
range thus observe the resulting \pair\ pair to reconstruct the
original photon properties. When the incident \gam\ photon is
polarized, the \pair\ azimuthal angles about the photon
direction are initially correlated with the polarization position
angle with a modulation factor $\sim$~20~\%, and the potential for
measuring polarization has long been recognized (Yang 1950, Wick 1950,
Maximon and Olsen 1962). However, as pointed out first by Kel'ner \et\
(1975), multiple scattering in the converting material and measurement
errors {\em exponentially} suppress the final modulation factor. This
is to be contrasted with detector angular resolution where the
dependence on scattering and errors is linear. Instruments optimized
to collect large numbers of photons with reasonable angular resolution
will thus not be adequate to measure polarization.

For this reason, initial hopes of sensitivity to polarization for
\COSB\ and \EGRET\ (\eg\ Kotov 1988, Caraveo \et\ 1988) have not been
confirmed by more detailed simulations and the analysis of actual
event data (Mattox 1992, Mattox \et\ 1990). Since then, the next
generation \gam\ instrument \GLAST\ has been proposed and
sophisticated simulation software developed to analyze its future
performance. Sufficient detail is included in the software model to
confidently predict polarization sensitivity. As the \GLAST\ design
has not been finalized at this stage, quantitative studies such as the
present one should be helpful in optimizing its performance for a
variety of physics goals, including polarization measurements.

We summarize the principal features of the pair production cross
section (\S\ref{PP}) and current \gam\ instruments (\S\ref{Current}).
Simple estimates of the polarization modulation factors are derived
(\S\ref{Estimates}) and Monte Carlo simulations performed for a single
conversion foil (\S\ref{Single}). The details of particle reconstruction
in multiple plane detectors are discussed (\S\ref{GLAST}) along with
our simulation results for \GLAST\ (\S\ref{Results}).

\section{\label{PP} Pair Production Modulation Factors}

Polarization can be measured only for bright sources so that the source
position is well determined and the observed \pair\ pair events
are completely described by six parameters: the photon energy
$\omega_\gamma$, the energy split $\epsilon^+$ (or $\epsilon^-$)
between the positron and electron, and the four spherical angles about
the known photon direction and an arbitrary position angle on the sky
(from North through East is the usual convention): $\theta^+$,
$\theta^-$, $\phi^+$, and $\phi^-$.  The \pair\ fork will tend to lie
in a plane that includes both the photon direction and its linear
polarization vector, resulting in a modulation of the pair production
cross section of the form:
$$
    \sigma_{\rm PP} \; \propto \; 1 + \lamp \; \cos^2{ \psip }
$$
where $\psip$ is the azimuthal angle of the \pair\ plane and $\lamp
\sim 1/3$ at high energies and when all production events are included.
Some authors prefer to use a quadrupole signal of the form $1 + {\rm R}
\cos 2\psip$, where $\lamp = 2 {\rm R} / (1 - {\rm R})$, as the average
of cosine is zero. If this property is desired, one can also use
$1 + \Lamp \; (\cos^{2}{ \psip } - 1/2)$, where $\lamp = \Lamp / (1 -
\Lamp/2)$. For partially polarized photons, both R and $\Lamp$ (but
not $\lamp$) are proportional to the amount of polarization.

The cross section topology is of course much richer than this simple
integral shows, and ideally the information in the cross section
should be used more fully in order to increase the sensitivity to
polarization.  When one looks for instance only at events where the
energy of the photon is split equally between the electron and
positron, $\lamp$ increases to $2/3$ (averaged again over all other
quantities). At lower photon energies $\omega_\gamma$~$<$~100~MeV,
$\lamp$ also increases. For plots of $\lamp$ as a function of several
parameters see Kotov (1988).

In an approximate sense, it is useful to think of the physical process
of pair production as creating the \pair\ pair exactly in the plane of
polarization. The recoil momentum of the nucleus $q$ then stochastically
changes the azimuthal angle of the fork to random directions. Sets of
events for which the nucleus recoil momentum is small thus have a
larger value of $\Lamp$. Events with small $q$ are much more likely
since the cross section includes a $1/q^{4}$ term; competing against
this dependence is the vanishing phase space at zero momentum transfer
$q$ to the atom.

\begin{figure}[tb]
\psfig{file=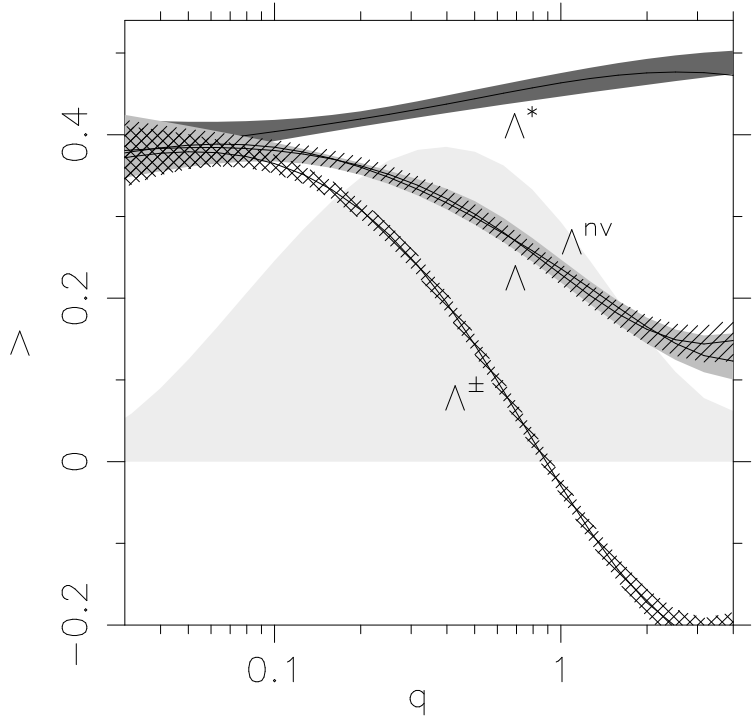,width=11cm}
\caption{Polarization signal at $\omega_\gamma =$~30~MeV as
a function of the nucleus recoil momentum~$q$. The top curve (darkest)
is the modulation amplitude $\Lamm$ of $\psim$, the middle curve is for
either $\Lamp$ of $\psip$ (lighter) or $\Lamv$ of $\psiv$ (hatched), and
the lower curve (cross-hatched) for $\Lambda$ of $\phi^+$ or $\phi^-$
(for which $\Lambda$ can be multiplied by $\sqrt{2}$ as there are two
possible measurements per event). Modulation is seen to vary from $\sim$
40~\% at zero momentum transfer, to negative values at large $q$ for
$\phi^\pm$. The light region shows the number of events as a function
of $q$ in arbitrary units. The curves are based on $10^7$ Monte Carlo
events with statistical 1$\sigma$ noise represented by the curve
widths. \label{recoil}}
\end{figure}

In constructing the azimuthal angle $\psip$ of the \pair\ fork plane,
the effect of the nucleus recoil momentum partially cancels. For this
reason, the polarization modulation $\Lamp$ of $\psip$ is larger than
that observed for the two azimuthal angles of the \pair\ tracks,
$\phi^+$ and $\phi^-$. Of course one expects the modulation to be
larger by a factor $\sqrt{2}$, as there are two possible measurements
in the case of $\phi^\pm$, and $\psip$ is some average of $\phi^\pm$.
The angle $\psip$ has, however, larger modulation than expected from
a simple average of the $\phi^\pm$. In fact, other weighted averages
of the two angles $\phi^+$ and $\phi^- \pm \; \pi$ are even more
successful in cancelling the coherent effects of the recoil momentum.
The best we have found is:
$$
    \psim \; \equiv \; w^+ \cdot \phi^+ + w^- \cdot (\phi^- \; \pm \; \pi),
    \;\;\; w^\pm = \frac{\sqrt{\epsilon^\pm \cdot \sin\theta^\pm}}{w^+ + w^-}
$$
where $\pi$ is added to or subtracted from $\phi^-$ depending on
whether $\phi^+$ is larger or smaller than $\phi^-$, respectively (and
with the branch cut [0,~$\pi$] for all angles). As seen in
Figure~\ref{recoil}, $\psim$ is almost entirely successful at
eliminating the dependence on recoil momentum.  Unfortunately, as
$\psim$ depends also on $\theta^\pm$ (as does $\psip$), it is more
difficult to measure than $\phi^\pm$, so there will be a tradeoff between
greater intrinsic modulation and increased measurement error. The
inclusion of $\epsilon^\pm$ in the definition of $\psim$ is not
essential and results in only a small improvement of the measured
modulation.

To determine the fork angles $\theta^\pm$ and $\phi^\pm$, a minimum of
three points is needed: the vertex of the pair production event and at
least one measurement along each of the \pair\ tracks. If the vertex
of the event is not measured, we are left with two points from which
only a single azimuthal angle $\psiv$ can be constructed. Since
scattering in each measurement plane significantly perturbs the \pair\
tracks, $\psiv$ may sometimes be the best estimator as it involves
only one measurement plane. As can naively be expected,
Figure~\ref{recoil} shows $\psiv$ to posses the same degree of
modulation as $\psip$.

\section{\label{Current} Current Detector Designs}

Pair production ceases to be the dominant conversion process below a
critical photon energy 610~MeV$/(Z + 1.24)$, where $Z$ is the atomic
number of the conversion material (in gases the corresponding formula
is 710~MeV$/(Z + 0.92)$). In addition, the resulting \pair\ pair must
posses enough energy to create two distinct tracks in the detector.
Both the \EGRET\ and \GLAST\ designs thus make use of many thin, high $Z$
conversion foils, with as little material as needed for the active
measurement planes and support structures sandwiched between them.

The \EGRET\ instrument has 27 tantalum pair production foils of
radiation thickness $\XX$~=~0.022 separated by 16.6~mm for a total of
about half a radiation length along 45~cm. In between the foils spark
chambers measure the track $x$ and $y$ coordinates with wires spaced 8~mm
apart. The lower portion of the detector has some additional spark
chambers with only small amounts of scattering material in between,
and two of these are rotated by $45^\circ$ to help in resolving the
ambiguity in pairing together $x$ and $y$ coordinates to form positions
in the detector. The \EGRET\ effective area decreases rapidly below
$\sim$~100~MeV and above a few GeV.

The current design of the \GLAST\ instrument calls for 12 lead
conversion foils of radiation thickness $\XX$~=~0.05 separated by 30~mm
for a total of about half a radiation length along 36~cm. Silicon strips
with $\sim$~250~$\mu$m pitch just below the foils measure the track
$x$ and $y$ coordinates. The amount of silicon in the strip planes is
not negligible and amounts to a total of 0.0077 radiation lengths for
the $x$ and $y$ planes, unless a single double-sided plane is used.

To limit the strip length, \GLAST\ is made of modules, with some area
lost between the modules. \GLAST\ accepts photons from almost 2$\pi$
of the sky, so that photons can have oblique incident angles to the
foils. This larger acceptance combined with better triggering results
in a much improved effective area for \GLAST\ as compared to \EGRET,
especially at low energies. At 30~MeV \GLAST\ should collect about 20
times more photons than \EGRET. Below 30~MeV, the effective area
remains much higher than that of earlier detectors, but drops rapidly
to a small fraction of the geometric area.

\section{\label{Estimates} Simple Estimates}

As noted in the introduction, the polarization modulation factors
are suppressed {\em exponentially} by measurement errors and multiple
scattering, as is easily shown by integrating the quadrupole in ``$\psi$''
(any $\psi$ or $\phi$) with a Gaussian error of width $\sigma_{\psi}$
(errors on $\psi$ are of course defined modulo $\pi/2$). The quantitative
decrease of ``$\Lamp$'' (again, any of the $\Lamp$'s) can then be understood
from some simple formulas. For an incident photon being measured a distance
$z$ from its conversion point, the error $\sigma_{\psi}$ due to multiple
scattering $\sigma_{\theta}$ and measurement error $\sigma_{xy}$ is
given by:
$$
    \sigma_{\psi} \; \approx \; \frac{\sigma(z)}{z \cdot \sin \theta },
    \;\;\; \sigma(z)^{2} \; \equiv \; \sigma_{xy}^{2} +
        \left(z \cdot \sigma_{\theta}(\XX, E) \right)^{2}
$$
A rough estimate of multiple scattering for $\XX$~=~$X/X_0$ radiation
lengths traversed at energy $E$ is given in Lynch and Dahl (1991):
$$
    \sigma_{\theta}(\XX, E) \; = \; \frac{13.6 \; {\rm MeV}}{E}
        \cdot \sqrt \XX \; ( 1 + 0.038 \cdot \log \XX )
$$
For an ensemble of events at some particular energy, $\theta$ can be
replaced with a characteristic polar angle $<\theta>$ for the set. The
median polar angles of the electron and positron vary strongly with
energy.  At 100~MeV, the median is $0.67^{\circ}$, and $2.2^{\circ}$
at 30~MeV.  In fact $<\theta> \; \propto 1/E$ is a good
approximation, where $E$ can be either the photon energy
$\omega_\gamma$, or the individual energies $\epsilon^\pm$ of the
\pair\ tracks. Inserting the expression for $\sigma(z)$ and assuming
small $<\theta>$ gives:
$$
    \sigma_{\psi}^{2} \; \approx \;
        \left( \frac{\sigma_{xy}}{z \, \cdot <\theta>} \right)^{2} +
        \left( \frac{\sigma_{\theta}}{<\theta>} \right)^{2}
$$

If the measurement error $\sigma_{xy}$ dominates the errors, low energies
are clearly favored as $<\theta>$ will be large (C is a constant):
$$
    \Lambda \; \propto \; \exp \left(- \; {\rm C} \cdot \left(
        \frac{\sigma_{xy}}{z \, \cdot <\theta>} \right)^{2} \right)
    \; \approx \; \exp \left(- \; {\rm C} \cdot \left(
        \frac{\sigma_{xy} \, \cdot \, E}{z} \right)^{2} \right)
$$

However, if multiple scattering dominates the errors, $\Lambda$ decreases
exponentially with $\XX$ and is independent of $<\theta>$ or $E$:
$$
    \Lambda \; \propto \; \exp \left(- \; {\rm C}
        \cdot \left( \frac{\sigma_{\theta}}{<\theta>} \right)^{2} \right)
    \; \approx \; \exp \left(- \; {\rm C} \cdot \XX \right)
$$
Photons can pair produce anywhere in the foil, so that in the
scattering dominated case the polarization signal will be due to
events that were produced close to the surface of the foil. The signal
will then only decrease linearly with $\XX$ (since the fraction of
events close to the surface will decrease linearly with thickness). We
can thus expect exponential decrease followed by linear decrease of
$\Lambda$ for large $\XX$. The last expression also reveals that there
is little benefit in choosing individual \pair\ tracks with a high
fraction of the photon energy in order to minimize the effects of
scattering, since these will have on average correspondingly small
polar angles (i.e. a plot of the $\psi$ error as a function of
$\epsilon^\pm$ is flat). Similarly, $\Lambda$ does not depend
strongly on $\omega_\gamma$. Since most astronomical objects
have a steep spectrum, polarization is again best measured at
low energies.

To summarize, in the energy range where errors are dominated by
scattering, all photons contribute equally to polarization
sensitivity, and at higher energies, polarization modulation decreases
exponentially. It does happen, however, that {\em particular} events
are very good, and others very bad. Equal energy split with large
opening angle is ideal. At the other extreme are events with no
measurable opening angle, and thus no azimuthal angle information. A
weighted histogram of the $\psip$ can thus be calculated in finding
$\Lambda$. It must be kept in mind though that assigning weights is
analogous to making cuts on the events, and so the statistical noise
is increased. The question then is whether there is sufficient gain in
the signal amplitude to offset this increased noise.

\section{\label{Single} Single Foil Results}

The preceding section showed that polarization is best measured at the
lowest possible photon energies. This explains the choice of
$\omega_\gamma$~=~30~MeV in Figure~\ref{recoil}, since the \GLAST\
effective area decreases rapidly at even lower energies. In this
energy range, measurement of the \pair\ track angles is clearly
dominated by multiple scattering.  Even though the \pair\ tracks will
often traverse many measurement planes, the later measurements will
have little memory of the initial \pair\ fork angles. Before
simulating a complete detector, it is thus useful to analyze the
situation when only one production foil is involved, and under the
assumption that the angles and energies of the \pair\ pair are
measured perfectly after leaving the conversion foil. These best case
results will present a benchmark for simulations of the complete
detector.

In our simulations we have used the polarized pair production cross
section $\sigma_{\rm PP}$ in the static field limit (no energy
transfer to the nucleus) and without screening, as given in \eg\ Jauch
\& Rohrlich (1975). The main effect of screening is to change the total cross
section which is not of interest here. We also neglect pair production
on the electron field, and the effects of the Coulomb field of the
\pair\ pair. We do not restrict ourselves to the limit of small
momentum transfer as Maximon and Olsen (1962) have done. For multiple
scattering we have used the expression of \S\ref{Estimates}.

\begin{figure}[ptbh]
\vspace{0.5cm}
\psfig{file=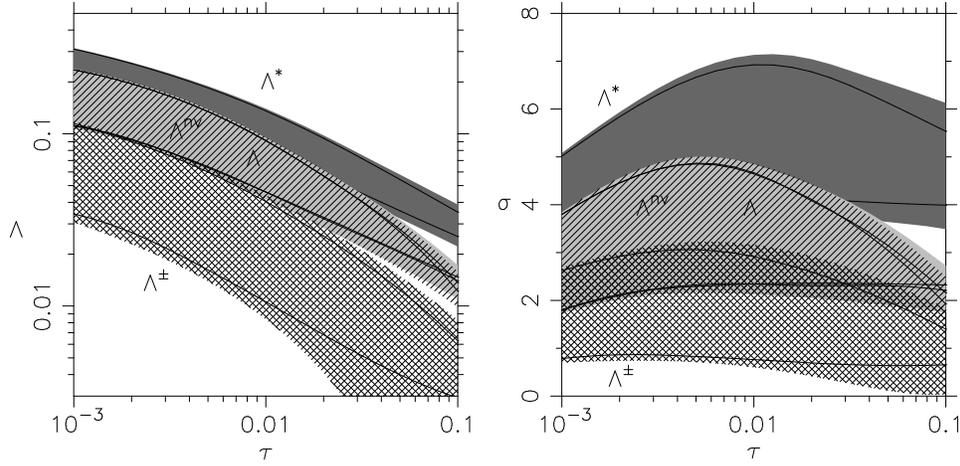,width=13.0cm}
\caption{Results at $\omega_\gamma =$~30~MeV for a single foil and no
measurement errors ($\sigma_{xy}$~=~0). The first panel shows the close
to exponential decrease of $\Lambda$ with the foil thickness $\XX$ in
radiation lengths. The curves are as in Figure~\ref{recoil}, with the
difference that the widths represent the loss in modulation expected from
$x$/$y$ pairing ambiguities in addition to the Monte Carlo 1$\sigma$
noise. The second panel shows the corresponding signal to noise ratio
in $\sigma$ for ($\XX/0.05$)~$\times$~$10^5$ photon conversions.
\label{single}}
\end{figure}

\begin{figure}[ptbh]
\psfig{file=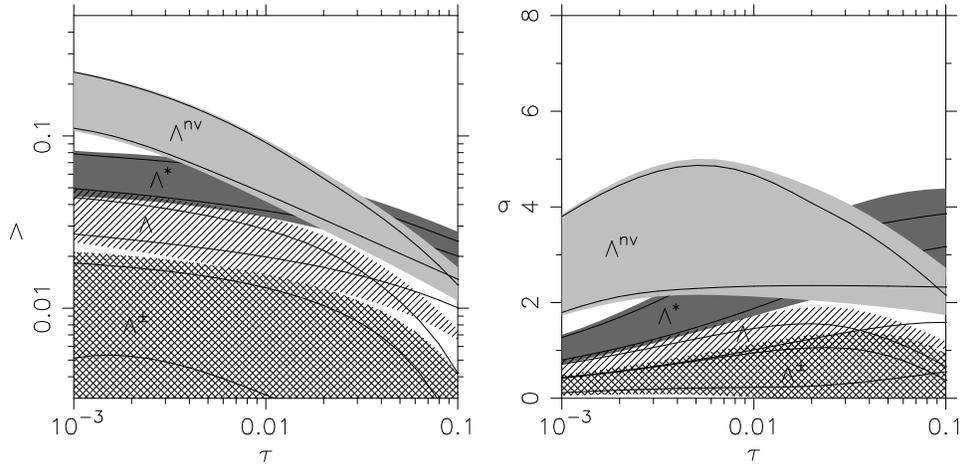,width=13.0cm}
\caption{Addition of an active layer just below the conversion foil
is needed to measure the event vertex. Scattering in this active layer
reduces the modulation as compared to Figure~\ref{single}. Here we have
assumed an active layer 0.0077 radiation lengths thick (\eg\ 700~$\mu$m
of silicon). Results for $\Lamv$ are copied from Figure~\ref{single} as
no vertex need be measured in constructing $\psiv$.
\label{single77}}
\end{figure}

The simulation results for the different $\Lambda$ are shown in Figure
\ref{single}. The first panel shows the almost exponential decrease of
the signal $\Lambda$ with increasing foil thickness $\XX$, and the second
panel the optimal choice of $\XX$ that maximizes the signal to noise
ratio for a single foil (or more generally for a fixed number of
foils). The larger $\Lambda$ at small $\XX$ is offset by the smaller
number of photons that will convert in a thinner foil. The curves shown
are for ($\XX/0.05$)~$\times$~$10^5$ events (i.e. constant flux, and $10^5$
conversions in a foil of thickness 0.05). The signal to noise ratio is
seen to be relatively flat between $\XX$~=~0.001 and 0.1. With no
measurement errors, a 3$\sigma$ observation would require only 8000
conversion events for foils with $\XX$~=~0.01 (using $\psim$). The figure
curves are smoothed versions of Monte Carlo sums of $5 \times 10^5$ events,
spaced every 0.1 in $\log{\XX}$ (so twenty runs for each curve).

Both \EGRET\ and \GLAST\ measure $x$ and $y$ coordinates of the \pair\
tracks in separate active layers, and the pairing together of the $x$
and $y$ to form detector positions is not known when more than one
track is recorded in the same layer. Resolving these ambiguities is
essential to the measurement of polarization, since the azimuthal
angles can change by $\sim$~$90^\circ$ when the wrong $x$ and $y$ are
combined. In Figure~\ref{single} the corresponding loss of modulation
is shown for the entire range from no ambiguity to full ambiguity
(i.e. from no errors, to bad pairing half of the time). With complete
ambiguity, 25000 events are needed for a 3$\sigma$ detection (factor
3 more).  As is explained in section \S\ref{Results}, some of this
loss can be regained by assigning event weights.

As mentioned in \S\ref{PP}, the event vertex must be measured in order
to determine all four track angles. An active layer is thus required
just below the conversion layer. A second active layer some distance
away will then measure the track angles from the vertex
position. Unfortunately, the tracks will also be scattered in the
first active layer as they must traverse its entire thickness. In the
case of \GLAST\, the active layers have significant thickness, 0.0077
radiation lengths in the current design (two 350~$\mu$m layers of
silicon). Figure~\ref{single77} shows the dramatic decrease in
modulation expected when scattering from the first active layer is
included. The effect is most pronounced for thin conversion foils. In
the case of $\psiv$ no vertex is needed, and so the curves have been
copied from Figure~\ref{single}. For most of the $\tau$ range, $\psiv$
now has larger modulation than $\psim$. For thinner active layers,
this will not be the case.

An important effect was not included in Figure~\ref{single77}. If the
active layer has significant thickness, additional photon conversions
will take place in the active layer itself.  These events will be
highly modulated, as the relevant $\tau$ is now the active layer
thickness, which is usually much less than the conversion foil
thickness. There are several complications though. If enough energy is
deposited in the active layer to allow measurement of the vertex, it
is not possible to distinguish such an event from a conversion in the
foils. In the contrary case, the event will not have a vertex and thus
only $\psiv$ can be constructed. When separate active planes are used
to measure the $x$ and $y$ coordinates, the event will have both
coordinates only half the time. Finally, as explained in
\S\ref{Current}, many conversions in the low $Z$ active material are
Compton events. Nevertheless, in the case of \GLAST, $\tau$~=~0.05
for the conversion foils and Figure~\ref{single77} applies, and
for the active layers, $\tau$~=~0.0077 and Figure~\ref{single}
applies.  Depending on the importance of these effects and $x/y$
ambiguities, photons that converted in the active layers may
dominate the polarization signal. This will be discussed further
in \S\ref{Results}.

In the second panels of Figures~\ref{single} and \ref{single77}, it
was assumed the resulting modulation is of the form $\cos^2 \psip$, so
that the signal to noise is given by $\Lambda/(0.028 \times \sqrt{N})$
for $N \times 10^4$ measurements of ``$\psi$'' (any $\psip$ or $\phi^\pm$).
We have found that measuring an asymmetry ratio results in a slightly
improved signal to noise as compared to fitting the $\cos^2 \psip$
modulation (approximately 10~\% larger). Clearly the measured
modulation is not described optimally by a $\cos^2$ and a better
estimator should be derived. This was not investigated further.

The results for single foils are also useful in interpreting designs
which are dominated by measurement errors and for which track angles
are effectively measured only after traversing {\em several} foils.
In this case the combined $\XX$ traversed is most relevant, and one
simply regains the situation where scattering dominates after $n$
foils have been traversed.  The foil thickness is then $\tau \times
n$. At 100~MeV, \EGRET\ is dominated by measurement errors with
$n$~$\sim$~4.

\section{\label{GLAST} Monte Carlo Simulation}

To simulate the \GLAST\ instrument in detail, we have used the
standard \GLAST\ Monte Carlo software which is an application built
upon Gismo (Atwood \et\ 1992). Gismo, written in C++, includes both
QED interactions adapted from the EGS4 code and hadronic processes
from the Gheisha code.  The \GLAST\ application includes many details
of the detector geometry that can affect the instrument performance,
for example, all materials with significant radiation length
(converter planes, silicon-strip detectors, readout electronics,
structural support elements, thermal control material), dead areas
around the edges of each module, and gaps between modules. Energy
deposition in the active portions of the silicon tracker and
calorimeter is also modeled in detail, and only the simulated
digital outputs are available to the event reconstruction classes.

We have modified the software in several important ways. First, the
polarized cross section as described in \S\ref{Single} was added as
a new interactor class. Second, the event reconstruction code was
entirely rewritten. This was necessary since the standard
reconstruction (simple linear fits through the tracker and
calorimeter) is inadequate for low energy events. The requirements for
polarization sensitivity are that individual \pair\ tracks and the
production vertex be identified correctly, giving estimates of the
track angles and energies from which the $\psip$, $\psim$ or $\psiv$
are constructed. Several approaches to event reconstruction were
experimented with, but we have had the most success with a brute
force search of all possible tracks through the silicon-strip
detector.

The silicon tracker output is a list of $x$ and $y$ strip ``hit''
coordinates at given $z$ heights. Our strategy is then to test all
possible sequences of hit positions ($x$,$y$,$z$), looking for
downward (increasing $z$) sequences with as little scatter as possible.
Writing the correct likelihood criterion for a track is a difficult
problem and it depends on the entire tracker output, so that we have
used an ad hoc expression instead:
\newcommand{\summ}{\textstyle \sum_i}
$$
    {\cal L} \; \equiv \; \summ v_i -
        \sqrt{ (n-2) \cdot \summ v_i^2 - (\summ v_i)^2 },
        \;\;\; v_i \; \equiv \; 1 - \alpha_i^2, \;\;
$$
where for a sequence of $n$ foils, we have ($n-2$) scattering angles
$\alpha_i$ (in radians). The second term of $\cal L$ favors tracks
for which the amount of scattering is constant along the track.
A list of likely (large~$\cal L$) tracks is built up during the
recursive search, and saved tracks are replaced with more likely
versions if they are judged to be similar enough (as defined by the
direction of the tracks and the number of points common to both).
We thus obtain a complete list of the best dissimilar tracks found.
Several million possible sequences must be tested in this way and
optimized algorithms from the GNU C++ library were used to allow
the search to finish within a second for most events (Sun UltraSparc).

As explained in section \S\ref{Single}, a complication is that the
pairing together of $x$ and~$y$ hit positions is not known when more
than one hit is recorded by the tracker at a given $z$. All possible
combinations are thus treated equally. This results in ``real'' tracks
also having ``shadow'' tracks present in the final list.  Resolving
these ambiguities is important to the measurement of polarization,
since we are interested in the track azimuthal angles.  If a track
crosses the boundaries of a module, the ambiguity is resolved
automatically. Otherwise the limited imaging capability of the
calorimeter must be combined with energy estimates from multiple
scattering in finding the most likely ``real'' tracks.  Individual
track energy estimates are thus an important element in finding likely
tracks. For our purposes a simple inversion of the Gaussian multiple
scattering formula was found to be sufficient.  With no measurement
errors, the expected energy resolution is only $1/\sqrt{n-2}$, where
$n$ is the number of foils traversed, so that for small $n$ the
estimate from scattering is useful only as a consistency check
against the energy deposited in the calorimeter.

A second processing stage then takes the saved tracks as input and
decides on the most likely combination of ``real'' tracks, using all
the information available. A score is calculated for each combination
that includes contributions due to: the ratio of energies estimated
from scattering and from the calorimeter, the fraction of the total
calorimeter energy assigned to tracks, the number of tracker hits
used, the number of ``shadow'' hit pairs used, and the existence of a
vertex. The highest score selects a particular set of tracks along
with estimates of their individual angles and energies. Which tracks
are ``real'' is of course most reliably determined close to the
calorimeter, and we must follow the tracks back to the vertex in order
to determine the original fork angles.  As \GLAST\ is entirely
scattering dominated at low energies, only the vertex and next layer
were used in calculating the \pair angles about the known incident
photon direction.  Otherwise, the next track positions would provide
additional estimates of the angles.

\section{\label{Results} Results and Discussion}

In general terms, this reconstruction algorithm delivers close to
optimal low energy response for \GLAST. All but 7~\% of the events
which convert within the tracker are reconstructed (some are too
complex to search in the allowed time or have no recognizable tracks),
individual tracks are identified, and their energies determined to
20~\%~FWHM or better at 30~MeV. Almost half of the events are found to
have two tracks with a well defined vertex. A single track is found
for 30~\% of the events; these correspond mainly to cases where the
\pair\ energy split was asymmetric. Three or more tracks are found in
the remaining 25~\% of cases. Compared to the standard \GLAST\
reconstruction, effective area is increased by more than 15~\% at
30~MeV. In addition, angular resolution is also improved, in part by
summing the track momenta to obtain the incident photon direction. The
single photon angular resolution is $1.8^\circ$ FWHM at 100~MeV, and
$6^\circ$ at 30~MeV.

\newcommand{\ww}[1]{\makebox[0.5cm][r]{#1}}

\begin{table}[p]

\label{Table}
\caption{Polarization sensitivity at 30~MeV, for both the simple and
detailed Monte Carlo sums. Single foil results are for Pb conversions,
$\sigma_{xy} = 0$, and no $x/y$ ambiguity. The required number of events~/
$10^3$ for a 3$\sigma$ detection is reported (100~\% polarized flux).
The number of events reflects both the modulation amplitude $\Lambda$
and fraction of all reconstructed events used. The required exposure
is thus given by the number of events~/ flux $\times$ effective area.
Results for $\cos^2{2\phi_0}$ event weights are also shown, where
$\phi_0$ is the angle between the detector $x$ axis and the polarization
direction. Each Monte Carlo sum was run for approximately 60 workstation
days, resulting in several million reconstructed events per run.}

\vspace{1. cm}
\begin{tabular}{p{1.1in}clclcl}

\multicolumn{6}{l}{\em Single foils} \\
& foil $\XX$ &
$\psip$ &
\makebox[0.1cm][c]{$\Lambda\times${\tiny 100}} &
\makebox[0.9cm][c]{$3\sigma$} & & \\
\hline
& & & & & & \\

Simple MC & 0.05 & $\psip$ &
2.70 & $ \ww{77}^{+25}_{-17}$ & & \\
& & $\psim$ &
5.41 & $ \ww{19}^{+3}_{-2}$ & & \\

Si added for vertex & 0.05 + 0.0077 & $\psip$ &
1.12 & $ \ww{448}^{+509}_{-189}$ & & \\
& & $\psim$ &
3.33 & $ \ww{51}^{+13}_{-9}$ & & \\

& & & & & & \\

Detailed MC & 0.05 & $\psip$ &
2.26 & $ \ww{140}^{+141}_{-56}$ & & \\
& & $\psim$ &
4.49 & $ \ww{36}^{+13}_{-9}$ & & \\

& 0.05 + 0.0077 & $\psip$ &
1.09 & $ \ww{622}^{+676}_{-258}$ & & \\
& & $\psim$ &
1.85 & $ \ww{216}^{+106}_{-61}$ & & \\

& 0.02 + 0.0038 & $\psip$ &
1.64 & $ \ww{258}^{+170}_{-86}$ & & \\
& & $\psim$ &
4.31 & $ \ww{37}^{+7}_{-6}$ & & \\
\hline

& & & & & & \\
\multicolumn{6}{l}{\em Multiple Planes} \\
& foil $\XX$ &
$\psip$ &
\makebox[0.1cm][c]{$\Lambda\times${\tiny 100}} &
\makebox[0.9cm][c]{$3\sigma$} &
\raisebox{-1ex}{\shortstack{\makebox[0.1cm][c]{$\Lambda\times${\tiny 100}} \\ $\cos^2$}} &
\makebox[0.9cm][c]{\raisebox{-1ex}{\shortstack{$3\sigma$ \\ $\cos^2$}}} \\
\hline
& & & & & & \\

Standard \GLAST\ & 0.05 + 0.0077 & $\psim$ &
0.90 & $ \ww{865}^{+318}_{-205}$ & 1.16 & $ \ww{725}^{+238}_{-160}$ \\
& & $\psiv$ &
0.89 & $ \ww{2200}^{+1519}_{-748}$ & 1.35 & $ \ww{1331}^{+647}_{-375}$ \\

No $x/y$ ambiguity & & $\psim$ &
0.96 & $ \ww{766}^{+276}_{-179}$ & & \\
& & $\psiv$ &
1.87 & $ \ww{499}^{+138}_{-98}$ & & \\

Double-sided Si & 0.05 + 0.0038 & $\psim$ &
0.87 & $ \ww{909}^{+444}_{-257}$ & 1.17 & $ \ww{700}^{+288}_{-178}$ \\

& & & & & & \\

Thin foils & 0.02 + 0.0077 & $\psim$ &
1.98 & $ \ww{162}^{+39}_{-29}$ & 2.17 & $ \ww{188}^{+50}_{-36}$ \\
& & $\psiv$ &
2.82 & $ \ww{235}^{+71}_{-49}$ & 4.10 & $ \ww{155}^{+37}_{-27}$ \\

Double-sided Si & 0.02 + 0.0038 & $\psim$ &
2.90 & $ \ww{75}^{+17}_{-13}$ & 3.22 & $ \ww{84}^{+21}_{-15}$ \\

\makebox[1.3cm][l]{Double-sided Si, no $x/y$} & & $\psim$ &
3.72 & $ \ww{45}^{+12}_{-9}$ & & \\
& & $\psiv$ &
7.11 & $ \ww{49}^{+14}_{-10}$ & & \\

& & & & & & \\

Rear Si & 0.05 + 0.0077 & $\psiv$ &
1.08 & $ \ww{586}^{+299}_{-170}$ & 1.79 & $ \ww{297}^{+98}_{-65}$ \\

& 0.02 + 0.0077 & $\psiv$ &
2.43 & $ \ww{129}^{+55}_{-34}$ & 3.37 & $ \ww{93}^{+33}_{-21}$ \\
\hline

\end{tabular}
\end{table}

The measured polarization sensitivity is, however, disappointing. From
Table~1, one finds that, without modifications, the \GLAST\ design would
require $7 \times 10^5$ photon events for a 3$\sigma$ detection of
100~\% polarized flux. With the new reconstruction algorithm, the \GLAST\
effective area is $\sim$~7000~cm$^2$ at 100~MeV, and $\sim$~5000~cm$^2$
at 30 MeV . For the Vela pulsar, \GLAST\ will thus collect $\sim$~6000
counts/day above 100~MeV, and $\sim$~8000 counts/day in the range 30--100~MeV.
A 3$\sigma$ detection would thus require two months of observing time.
Observations of $\sim$~20~\% polarization in several phase bins would not
be possible. Changes to the \GLAST\ design in order to improve
polarization sensitivity are clearly indicated.

In order to verify that the \GLAST\ sums agree with the simple
simulation described in \S\ref{Single}, we simulated a single
conversion foil with the \GLAST\ code, using only two active layers:
one directly below the conversion foil, and one at a large
distance. The resolution of the active layers was arbitrarily good,
and there was no $x/y$ ambiguity.  Table~1 shows the \GLAST\ results
to be comparable to the simple simulation, \eg\ $\Lamm$ of 0.0449
compared to 0.0541. The simple form for multiple scattering (and the
absence of Bhabha and Moller scattering) adopted in the simulations of
\S\ref{Single} can easily account for these differences, since all
errors affect $\lambda$ exponentially. The Monte Carlo sums of Table~1
were all allowed to run for approximately 60 workstation days,
resulting in several million events per configuration. Monte Carlo
noise is then sufficiently small to make meaningful measurements of
all ``$\Lambda$'' except $\Lambda^\pm$.

In going from a single conversion foil to the complete \GLAST\
geometry, there is a further decrease in modulation, \eg\ from 0.0185
to 0.0090 for $\Lamm$. This decrease must be due to a combination of
$x/y$ ambiguities and other reconstruction errors. Using our simple
simulations as a guide, we expect the modulation to decrease by half
when $x/y$ ambiguities are present. Although the reconstruction
algorithm correctly selects ``real'' tracks over ``shadow'' tracks,
near the vertex we find that the position ambiguities are not well
resolved since both ``real'' and ``shadow'' positions are closely
clustered together. The error made by pairing the $x$ and~$y$
coordinates incorrectly can be very different depending on the
orientation of the instrument axes as compared to the original track
azimuthal angle. The best situation is when the fork plane contains
one of the axes so that there is no ambiguity, and the worst is when
the axes are rotated by $45^\circ$ from this position so that an
incorrect pairing will give an error of $90^\circ$.  The orientation
of the \pair fork plane before scattering is of course not known, but
the events can be weighted with $\cos^2{2\phi_0}$, where $\phi_0$ is
the angle between the detector $x$ axis and the assumed polarization
direction. The modulation is increased in this way, \eg\ from 0.0090
to 0.0116 for $\Lamm$, and results in increased sensitivity to
polarization.

The $x/y$ ambiguities can be resolved directly by modifications to the
\GLAST\ design. For instance, if at least one strip layer per module
is rotated by an angle $0^\circ$--$45^\circ$, $x/y$ ambiguities can
in principle be resolved. This complicates the module design as the
strips are then of unequal length, and the readout electronics must
also be mounted differently. The same result can also be achieved by
correlating energy deposition in the $x$ and $y$ coordinates, or, for
the last active layer, by adopting a calorimeter with better imaging
capabilities, such as the proposed ``Sci-Fi'' design.  The modulation
signal from the last foil may approach in this case the optimal single
foil value reported in Table~1, since both angles and individual track
energies would be reliably measured. We have modified the \GLAST\ code
to eliminate $x/y$ ambiguities. Results as given in Table~1. The
effect is most dramatic for $\psiv$, where modulation is observed to
more than double.

Returning to the standard \GLAST\ design, the table also shows $\Lamv$
to be of the same magnitude as $\Lamm$.  Unfortunately, in the initial
layer, separate tracks can be resolved for very few events, as there
is no separation between the conversion foil and first active layer.
As a result, too few events are resolved to be useful for
polarization.  A simple modification of the \GLAST\ design would be to
separate the active layers from the conversion foils, contrary to the
current design. The results for this modification are noted in the
table with ``Rear foils''.  Rear foils are seen to improve the
sensitivity to polarization by 50~\%.  Angular resolution at low
energies is decreased though, from $1.8^\circ$ to $2.6^\circ$ FWHM
at 100~MeV; there is no such effect at high energies, as errors are
dominated by the strip pitch.

A more promising modification of the \GLAST\ design is to increase the
number of conversion foils while decreasing their thickness, as now
seems possible. The simple simulation of Figure~\ref{single} shows
that halving the foil thickness should increase modulation by at least
50~\%. We have simulated the case $\tau$~=~0.02. Polarization
modulation more than doubles, as can be seen from Table~1. A less
drastic modification of the \GLAST\ design would be to make the active
Si layers thinner, since the scattering in the first active layer has
a large effect on the modulation of $\psim$.  Results for half the
usual Si thickness (double-sided Si) are given in Table~1 and show
that the modulation does not improve over the standard design. This
surprising result can be explained from the arguments of \S\ref{Single}.
For $\tau$~=~0.05 foils, Si events dominate the polarization sensitivity,
and half of these events are lost with the decreased Si thickness.

Doubled-sided Si layers, i.e. with the $x$ strips on one side, and the
$y$ strips on the other side of the same Si layer, are nevertheless
desirable. When two separate layers are used, half of the events which
convert in the Si will convert in the second layer and will thus have
only one coordinate. These events are not useful for polarization
measurements (although one then knows that the conversion occured in
the Si and these events can thus be singled out). Double-sided Si is
thus superior to two single-sided Si layers of half thickness, as all
Si events will have both coordinates. In essence, this explains why
the modulation does not changed when the two Si planes are replaced
with one double-side plane of same thickness. The same amount of
useful Si events will be collected in both cases. If the conversion
foils are made thinner, however, Si events cease to dominate the
polarization signal, and double-sided Si planes do improve the
modulation, by almost 50~\% when $\tau$~=~0.02. As discussed
previously, the critical energy where the Compton cross section is
equal to the pair production cross section is proportional to 1/Z, and
for Si half the events are Compton events at 12 MeV; at energies lower
than 30~MeV, Si events are thus contaminated by Compton events.

What do these results mean for observations of polarized pulsar
emission? If no model is assumed, the observed \gam\ pulse must be
divided into several phase bins to map the position angle variations
as a function of phase.  Simple models of the \gam\ pulsar emission
processes predict large polarizations of $\sim$~80~\%.  At the
opposite extreme, polarization may be as low as observed in the
optical, or 20~\%.  Adopting this pessimistic view, observations of
the polarization seep would be challenging.  Combining the realistic
improvements to \GLAST, thin $\tau$~=~0.02 conversion foils and
double-sided Si layers, three years of Vela data would be needed
for six such 3$\sigma$ measurements. Clearly, this is at the limit
of what is possible. On the other hand, 80~\% polarized flux
from Vela could be detected with 3$\sigma$ significance in two
weeks. At a minimum, \GLAST\ will thus be able to rule out models that
predict a high degree of linear polarization.  Furthermore, additional
improvements to \GLAST\ are possible. With no $x/y$ ambiguities for
instance, the required exposure is reduced by almost half (for
$\tau$~=~0.02).  More than a factor two difference remains between the
single foil results and those for multiple planes; better event
reconstruction can reduce this gap. It may also be possible to collect
photons of energies somewhat lower than 30~MeV, i.e. in the range
10--30~MeV.  Extending the sensitive range of \GLAST\ to 10~MeV would
essentially double the number of photons collected. Finally, Monte
Carlo sums above 30~MeV show moderately improved polarization
sensitivity (per photon), so the required exposures may be
somewhat shorter.

Polarization measurements are at the opposite extreme to angular
resolution in terms of optimizing the tradeoff between detector errors
and the number of photons collected. Other physics goals such as the
imaging of extended or confused regions have intermediate requirements.
Hopefully at this early stage in the design of \GLAST\, a suitable
optimization of the instrument can be found that will satisfy all the
\GLAST\ physics goals. Although challenging, polarization measurements
{\em are possible} with an improved \GLAST. More extensive simulations
will clearly be necessary in finding optimal solutions to both the
hardware design and event reconstruction issues.

\acknowledgements
This work was supported in part by NASA grants NAGW 2963, NAG 5-2037
and NAG 5-3231. Helpful discussions with Roger W. Romani, Bill Tompkins,
T. Burnett and W.B. Atwood are acknowledged. The software and data
are available from http://astro.stanford.edu/home/ion.

\clearpage
\section*{REFERENCES}
\begin{description}
\item Atwood, W.B., \et\ 1992, Int. J. of Mod. Phys., C3, 459
\item Caraveo, P.A., \et\ 1988, ApJ, 327, 203
\item Jauch, J.M. and Rohrlich, F. 1975, Springer-Verlag (New York)
\item Kel'ner, S.R. \et\ 1975, Yad. Fiz., 21, 604
\item Kotov, Yu.D. 1988, Space~Sci.~Rev., 49, 185
\item Lynch, G.R. and Dahl, O.I. 1991, Nucl. Instr. and Meth., B58, 6
\item Mattox, J.R. 1991, Exp. Astron., 2, 75
\item Mattox, J.R., \et\ 1990, ApJ, 363, 270
\item Maximon, L.C. and Olsen, H. 1962, Phys. Rev., 126, 310
\item Romani, R.W. and Yadigaroglu, I.-A. 1995, ApJ, 438, 314
\item Wick, G.C. 1950, Phys. Rev., 81, 467
\item Yang, C.N. 1950, Phys. Rev., 77, 722
\end{description}
\end{document}